\UseRawInputEncoding
\documentclass[12pt]{article}
\pdfoutput=1 
\usepackage{amssymb, amsthm,amsmath,amsfonts}
\usepackage{graphicx}
\usepackage{xcolor, colortbl}
\usepackage{empheq}
\usepackage[pdftex, bookmarks=true,colorlinks,linkcolor=red,urlcolor=blue, citecolor=blue]{hyperref}

\usepackage{amsthm}
\usepackage{array}
\usepackage{soul}
\usepackage{float}
\usepackage{bm}

\makeatletter\@addtoreset{equation}{section}\makeatother

\newcommand{\preprint}[1]{\begin{table}[t]  
             \begin{flushright}               
             {#1}                             
             \end{flushright}                 
             \end{table}}                     
\renewcommand{\title}[1]{\vbox{\center\LARGE{#1}}\vspace{5mm}}
\renewcommand{\author}[1]{\vbox{\center#1}\vspace{5mm}}
\newcommand{\address}[1]{\vbox{\center\em#1}}

\def\be{\begin{eqnarray}}
\def\ee{\end{eqnarray}}
\def\bea{\begin{eqnarray}}
\def\eea{\end{eqnarray}}

\def\Dslash{\,\,{\raise.15ex\hbox{/}\mkern-12mu D}}
\def\Dbarslash{\,\,{\raise.15ex\hbox{/}\mkern-12mu {\bar D}}}
\def\delslash{\,\,{\raise.15ex\hbox{/}\mkern-9mu \partial}}
\def\delbarslash{\,\,{\raise.15ex\hbox{/}\mkern-9mu {\bar\partial}}}
\def\pslash{\,\,{\raise.15ex\hbox{/}\mkern-9mu p}}
\def\calDslash{\,\,{\raise.15ex\hbox{/}\mkern-12mu {\cal D}}}

\def\lae{\mathrel{\mathop{\smash{\lower .5 ex \hbox{$\stackrel<\sim$}}}}}
\def\lae{\mathrel{\mathop{\smash{\lower .5 ex \hbox{$\stackrel>\sim$}}}}}

\def\arXiv#1{\href{http://arxiv.org/abs/#1}{arXiv:#1}}
\def\arXiv#1#2{\href{http://arxiv.org/abs/#1}{arXiv:#1}}

\def\doi#1#2{\href{http://doi.org/#1}{#2}}

\newcommand{\pd}{\partial}

\begin{document}

\unitlength = .8mm

\begin{titlepage}
\vspace{.5cm}
\preprint{}

\begin{center}
\hfill \\
\hfill \\
\vskip 1cm

\title{\boldmath 
Black hole singularities across phase transitions}
\vskip 0.5cm
{Yan Liu}\footnote{Email: {\tt yanliu@buaa.edu.cn}},
{Hong-Da Lyu}\footnote{Email: {\tt hongdalyu@buaa.edu.cn}}
and
{Avinash Raju}\footnote{Email: {\tt avinashraju777@gmail.com}}
\address{Center for Gravitational Physics, Department of Space Science, \\ and International Research Institute
of Multidisciplinary Science,
\\ Beihang University,  Beijing 100191, China}
\end{center}
\vskip 1.5cm

\abstract{
We study the behavior of black hole singularities across the Hawking-Page phase transitions, uncovering possible connections between the physics inside and outside the horizon.  We focus on the case of spacelike singularities in Einstein-scalar theory which are of the Kasner form. We find that the Kasner exponents are continuous and non-differentiable during the second order phase transitions, while discontinuous in the first order phase transitions. We give some arguments on the universality of this behavior. We also discuss possible observables in the dual field theory which encode the Kasner exponents.
}

\end{titlepage}

\begingroup 
\hypersetup{linkcolor=black}
\tableofcontents
\endgroup

\section{Introduction}

Black holes are robust predictions of Einstein's gravitational theory, which have  attracted immense attention from both the theoretical community and experimental observations. However, the physics of black hole interior is still mysterious. It is  believed that the appearance of singularities behind the black hole horizon are unavoidable \cite{Hawking:1969sw}. The description of the mathematical structure of black hole singularities has been studied by Belinski, Khalatnikov and Lifshitz (BKL) \cite{BKL}. 

Among the families of spacelike BKL singularities, 
the most well-known singularities are the ones in Schwarzschild black hole and FLRW cosmology. More generally, in the simplest homogeneous cases, the BKL singularities are known as Kasner singularities  \cite{Kasner:1921zz, BKL-book} which take the form  
characterized by Kasner exponents. Recently it was also shown that the geometry near the singularity inside the black hole solutions of a class of Einstein-scalar theories in AdS is of Kasner form \cite{Frenkel:2020ysx}.\footnote{Other work on the Kasner singularities in gravitational systems can be found in e.g. \cite{Hartnoll:2020rwq,Hartnoll:2020fhc,Cai:2020wrp,Wang:2020nkd,Yang:2021civ, Mansoori:2021wxf, Shaghoulian:2016umj}. Previous studies on singularities in asymptotic flat spacetime in Einstein-scalar gravity can be found in \cite{bkl-conjecture}.  } More explicitly, in presence of scalar field, close to the Kasner singularity, the $d+1$ dimensional geometry and the scalar field are of the form
\be\label{eq:kasner}
ds^2\sim -d\tau^2+\tau^{2p_t} dt^2+\tau^{2p_i}dx^i dx^i\,,~~~\phi \sim -\sqrt{2}p_\phi \log\tau
\ee
where $\tau$ is a function of radial coordinate and 
the Kasner exponents $p_t, p_i, p_\phi$ satisfy $p_t+\sum_{i=1}^{d-1} p_i=p_t^2+\sum_{i=1}^{d-1}p_i^2+p_\phi^2=1$. In general the Kasner metric has curvature singularity at $\tau=0$ except the case where one of the Kasner exponents $p_t, p_i$ is $1$ while the other Kasner exponents vanishes. 

Obviously the BKL picture is based on the classical gravitational theory which would break down close to the singularity. It is expected that the singularity will be resolved in the full quantum theory of gravity. Nevertheless, we believe that analyzing the singularity within the framework of classical gravity is still  important and might shed light on the singularity  resolution mechanism. Here we shall discuss the singularity using classical gravity and focus on the black hole systems with the singularities of Kasner form.\footnote{For the generic case, for example we could consider inhomogenities, when we approach the singularity, the Kasner exponent oscillates \cite{BKL-book, Damour:2002et} and we will not consider this case. }  
  

The goal of this paper is to study the behavior of the black hole singularities during the black hole phase transitions. It is well-known that black hole phase transitions could be studied within framework of black hole thermodynamics from the approach of Euclidean gravity which is related to the physics outside of horizon. Therefore our study is expected to show a further link between the physics inside the black hole horizon and that of the outside.  We shall study a four dimensional Einstein-scalar theories in AdS with double trace deformations. By tuning the ratio between temperature and the double trace deformation parameter (which is the only dimensionless tunable parameter), the system undergoes different orders of phase transitions by choosing different scalar potentials.  
Since the black holes solutions have spacelike Kasner singularities inside the horizon, we shall study the behaviors of the Kasner exponents during the phase transitions.

Although the black hole singularities are located behind the horizon of the black holes, their information could  can be extracted from the physical quantities of the dual field theory in the context of AdS/CFT
correspondence, e.g. correlation functions  \cite{Fidkowski:2003nf, Festuccia:2005pi,Grinberg:2020fdj},  entanglement entropies \cite{Hartman:2013qma} etc. 
The fact that the information about singularity can be extracted from the dual boundary field theory is consistent with the black hole complementarity \cite{Susskind:1993if}.
We will discuss how to probe the Kasner exponents of the black holes from spacelike and timelike geodesics.

Our paper is organized as follows. In Sec. \ref{sec:setup}, we set-up the holographic model and present all the necessary ingredients for the numerical calculations to study the phase transitions and black hole singularities. In Sec. \ref{sec:pt} we discuss the numerical results on the behaviors of singularities across the phase transitions. We conclude and discuss some open questions in Sec. \ref{sec:condis}.

\section{Set-up}
\label{sec:setup}
We begin by collecting all the necessary ingredients for constructing black hole solutions and  studying the phase transitions in a four dimensional Einstein scalar theory in AdS. We will first present the four dimensional Einstein-scalar theory under consideration and show the numerical strategy to solve the system from the boundary to the singularity. Then we will study the thermodynamics of the black hole solutions in order to study the phase diagrams of the system from which the singularity behavior could be uncovered.

\subsection{Four dimensional Einstein-scalar theory}

We consider a 3+1 dimensional theory of real scalar field coupled to Einstein's gravity\footnote{We have set the AdS radius $L=1$.}
\begin{eqnarray}
\label{EinScalarAction}
S = \frac{1}{16\pi G} \int d^4x\;\sqrt{-g}\,\Big[R+6 - (\nabla \phi)^2 - V(\phi)\Big]
\end{eqnarray}
with a potential $V$ for the scalar field given by
\begin{eqnarray}
\label{eq:potential}
V(\phi) =  m^2 \phi^2 - \lambda_3\, \phi^3 + \lambda_4\,\phi^4\,.
\end{eqnarray} 
To have a positive $V$ when $\phi\to\pm\infty$, we should have $\lambda_4>0$. Note that when $\lambda_3\neq 0$, the $Z_2$ symmetry of the system $\phi\to-\phi$ is broken, and the potential $V(\phi)$ has a W-shape with two different local minima values located at 
$\phi_{1,2}=\frac{3\lambda_3\pm\sqrt{9\lambda_3^2-32 m^2\lambda_4}}{8\lambda_4}$. We shall focus on the case $\lambda_3\geq 0$. We will be interested in a class of solutions that asymptotically approach AdS$_4$ at the boundary. 
According to the AdS/CFT dictionary, the boundary CFT has a dual operator whose scaling dimension is determined by the mass of the bulk scalar field.  
 
We are interested in the finite temperature phases of the CFT, which are dual to black holes in the bulk. We follow \cite{Frenkel:2020ysx} to make following ansatz for the asymptotically AdS$_4$ metric and scalar field
\begin{eqnarray}
\label{field-ansatz0}
	ds^2 = \frac{1}{r^2} \left(-f(r)e^{-\chi(r)}dt^2 +\frac{dr^2}{f(r)} + dx^2 + dy^2\right)\,,~~~\quad \phi = \phi(r)\,
\end{eqnarray}
where $f$, $\chi$ are only functions of $r$. The AdS$_4$ boundary is located at $r\to 0$ while the singularity is located at $r\to\infty$. It is also useful to work in the Eddington-Finkelstein coordinate where the above metric has the following form 
\begin{eqnarray}\label{field-ansatz}
	ds^2 = \frac{1}{r^2} \left(-f(r)e^{-\chi(r)}dv^2 +2e^{-\chi(r)/2} dv dr + dx^2 + dy^2\right)\,,~~~\quad \phi = \phi(r)\,
\end{eqnarray}
where $v$ is the infalling Eddington-Finkelstein coordinate. Obviously the planar Schwarzschild black hole is a solution of the form (\ref{field-ansatz}) with $f=1-\frac{r^3}{r_h^3}, ~\chi=\phi=0$. We will be interested in the black hole solution with nontrivial scalar hair. 

We focus on the case with $m^2=-2$. The equations of motion are given by
\bea
\label{eq:HoloEOM}
	\begin{split}
		\chi' - r\phi'^2 &= 0\,, \\
		f'-\Big(\frac{r \phi'^2}{2}+\frac{3}{r}\Big)\, f+\frac{3}{r}-\frac{V}{2r}&= 0\,,\\
		\phi'' + \Big(\frac{ f'}{f}-\frac{2}{r}-\frac{\chi'}{2} \Big)\phi'  - \frac{\pd_{\phi} V}{2r^2f} &= 0\,,
	\end{split}
\eea
where the prime is the derivative with respect to the radial coordinate $r$.

From \eqref{eq:HoloEOM} we know that near the asymptotic AdS$_4$ boundary the fields behave as a power series in $r$ as 
\bea
\label{eq:nbexp}
\begin{split}
	f&= 1 + \frac{\alpha^2}{2}r^2 + \left(m_T  - 2\alpha^3 \lambda_3 \log r \right) r^3+ \cdots
		 \\
		\chi&= \chi_0+ \frac{\alpha^2}{2}r^2 + \frac{1}{3} \left(4\alpha \beta- \alpha^3 \lambda_3-6\alpha^3 \lambda_3 \log r \right) r^3+ \cdots \\
		\phi &=  \alpha r+  \left(\beta-\frac{3}{2}\alpha^2 \lambda_3 \log r \right) r^2+ \frac{1}{8}\left(-12\alpha\beta \lambda_3 + \alpha^3(2-27\lambda_3^2 + 8\lambda_4) + 18\alpha^3 \lambda_3^2\, \log r  \right) r^3+ \cdots 
\end{split}
\eea
The scalar field $\phi$ in the bulk is dual to a scalar operator $\mathcal{O}$. We can interpret either the value $\alpha$ or $\beta$ as the VEV of $\mathcal{O}$ for the current choice of mass parameter here, which are known as alternative quantization and standard quantization respectively \cite{Klebanov:1999tb}. 

Here we use the alternative quantization for the scalar field, i.e. we set $\alpha$ as the expectation value of the dual operator. Note that the  conformal dimension of the operator is one. In this case, we can consider a relevant deformation of the boundary field theory by a double-trace operator \cite{Witten:2001ua,dtd2}
\begin{eqnarray}
S \rightarrow S - \kappa \int d^3x \;\mathcal{O}^2\,.
\end{eqnarray}
When $\kappa<0$, this deformation makes the dual system easier to condensate and might induce a phase transition at finite temperature \cite{Faulkner:2010gj}. Therefore we shall focus on the parameter regime with $\kappa<0$.   
With double trace deformations, now the source of the operator in the boundary is $ \kappa \alpha - \beta$ while $\alpha$ is the expectation value of the operator \cite{Witten:2001ua, Papadimitriou:2007sj}. 

Holography with double trace deformation
has been widely studied in the literature to explore the physics of phase transitions in the context of AdS/CMT, see e.g.   \cite{Faulkner:2010gj,She:2011cm,  Iqbal:2011aj,Mefford:2014gia}. 
In case of the symmetric potential in \eqref{EinScalarAction}, \textit{i.e.} $\lambda_3=0$ and $\kappa<0$, there is a critical temperature $T_c$ below which there is a nonzero value for $\langle \mathcal{O}\rangle$  \cite{Faulkner:2010gj,Mefford:2014gia}. In the following we shall consider the general cases of the potential and study the behavior of the phase transitions in the dual system as well as the behaviors of the singularities during the phase transition in the bulk. 

At finite temperature, near horizon the fields can be expanded as  
\begin{align}
\begin{split}\label{eq:nhexpansion}
		f &= -\frac{V(\phi_h)-6}{2\,r_h} (r_h-r)  + \cdots \\ 
		\chi&= \chi_h+ \frac{\phi_h^2 (4+3\lambda_3 \phi_h -4\lambda_4 \phi_h^2)^2}{r_h(V(\phi_h)-6)} (r_h-r) + \cdots \\ 
		\phi &= \phi_h + \frac{\phi_h (4+3\lambda_3 \phi_h -4\lambda_4 \phi_h^2)^2}{r_h(V(\phi_h)-6)}  (r_h-r) + \cdots
	\end{split}
\end{align}
where $V(\phi_h)=-2\phi_h^2-\lambda_3\phi_h^3+\lambda_4\phi_h^4$. The zero temperature near horizon condition will be discussed in the next section. 

There are three free parameters $\chi_h, r_h$ and $\phi_h$ at the horizon. The Hawking temperature of the black hole is determined by\footnote{Note that to get the temperature, we should set $\chi_0$ in \eqref{eq:nbexp} to be zero.} 
\begin{align}
       T= \frac{|f'(r_h)| e^{-\chi_h/2}}{4\pi}
       =\frac{(6-V(\phi_h)) e^{-\chi_h/2} }{8\pi \,r_h}\,,
\end{align}
from which we have the relation between $\phi_h$ and temperature $T$ 
\begin{eqnarray}
	8\pi r_h T e^{\chi_h/2} + V(\phi_h) -6= 0\,.
\end{eqnarray}
Given the fact that the temperature should be positive, the above relation puts a constraint on the allowed $\phi_h$.
For the geometry to flow to asymptotic AdS$_4$ with appropriate AdS radius,
$\phi_h$ should satisfy 
$\phi_1 \leq \phi_h \leq \phi_2$, where $\phi_{1,2}$ are two minima of the potential for the scalar field.

The following two scaling symmetries are useful for solving the system. The first one is 
\begin{align}\label{scalsym1}
&v\rightarrow b v\,,~~\chi\rightarrow \chi+2 \log b\,,
\end{align}
and the second one is 
\begin{align}\label{scalsym2}
&(v,r,x,y) \rightarrow b(v,r,x,y)\,.
\end{align}
We can set $\chi_h=0$ using (\ref{scalsym1}) and set $r_h=1$ using  (\ref{scalsym2}), then near the horizon we only have one free parameter $\phi_h$. We can take the $\phi_h$ as a shooting parameter and integrate the equations from horizon to AdS$_4$ boundary numerically to obtain the black hole solution and therefore we can get $\{m_T, \chi_0, \alpha,\beta\}$ near AdS$_4$ boundary. From the numerical solution we obtained, $\chi_0$ can be scaled to be zero using the scaling symmetry (\ref{scalsym1}) and then we have a nonzero $\chi_h$ near the horizon. By imposing that the dual theory is sourceless with the double trace deformation, i.e. $\kappa\alpha-\beta=0$, we obtain a nonzero $\kappa$. Therefore, for a given $\phi_h$, we have the only dimensionless parameter $T/(-\kappa)$ in dual field theory. 
Note that when $\lambda_3\neq 0$, the logarithmic term in \eqref{eq:nbexp}  indicates a scaling anomaly for the double trace deformation parameter.\footnote{A similar scaling anomaly of the marginal double trace  deformation parameter was found in a different context \cite{Hofman:2017vwr,Grozdanov:2017kyl}. One difference is that here the double trace deformation parameter is relevant.}
One could introduce an explicit UV scale by replacing $\log r$ in \eqref{eq:nbexp} as $\log (\Lambda r)$. With a specific scale $\Lambda$ which we set to be 1, and we also fix the boundary value $\kappa=-1$ in this case. Thus for $\lambda_3\neq 0$ the theories under consideration are defined at the same  renormalization scale with $\kappa=-1$.

It is necessary to know the interior region of the black hole solution to study the behavior of Kasner exponents during black hole phase transition. So we need to do the numerical integration from horizon to a large enough $r_s$ that can be thought as the singularity.  
From the equations of the system \eqref{eq:HoloEOM}, when $r \rightarrow \infty$, the solutions behave as 
\begin{align}
\label{eq:rel1}
\phi = \sqrt{2}c \log r+ \cdots,~~~ \chi = 2c^2 \log r + \chi_1 + \cdots,~~~ f = -f_1 r^{3+c^2} + \cdots\,,
\end{align}
where $f_1>0$, and $c$ is a constant. This behavior is the same as \cite{Hartnoll:2020rwq} because the higher power term of $\phi$ in $V$ is subleading near the singularity. The values of $f_1$ and $c$  depend on the specific black hole solution. For Schwarzschild black hole we have $c=0$.  The near singularity solution can be changed to Kasner form (\ref{eq:kasner}) by coordinate transformation $r \rightarrow 1/(\tau^{\frac{2}{3+c^2}})$ and the Kasner exponents are related to  $c$ as 
\begin{align}
\label{eq:rel2}
p_x=p_y=\frac{2}{3+c^2}\,,~~~ p_t=\frac{c^2-1}{3+c^2}\,,~~~ p_\phi=\frac{2\sqrt 2 c}{3+c^2}\,.
\end{align}
Note that here we focus on the cases with isotropic and homogeneous spatial directions in the dual theory and we do not see any oscillatory behavior of the Kasner exponents in the examples considered.\footnote{One example where the Kasner exponents oscillate is for holographic superconductor system \cite{Hartnoll:2020fhc, Cai:2020wrp}. More generally, if we consider the spatial dependence of background fields, usually the Kasner exponents show chaotic behavior \cite{BKL-book}. } Nevertheless it would be extremely interesting to study the behavior of singularity during the  dynamical phase transition, e.g. \cite{Janik:2017ykj}. 

\subsubsection{Radially conserved charge}

The system is invariant under the transformation \eqref{scalsym1}. 
There is a radially conserved charge associated to this symmetry which is given by 
\begin{eqnarray}\label{conserved-charge}
	Q &= \frac{1}{r^2}\, e^{-\frac{\chi}{2}}\,\left(f'- f\chi'\right)\,,
\end{eqnarray}
satisfying $Q' = 0$. The conserved charge associated to the scaling symmetry (\ref{scalsym2}) can also be computed from the Noether theorem and using the equations of motion \eqref{eq:HoloEOM} one can show that this conserved charge is exactly the same as \eqref{conserved-charge}. The conserved charge \eqref{conserved-charge} can be used to check the numerical code.

\begin{figure}[h!]
\begin{center}
\includegraphics[
width=0.54\textwidth]{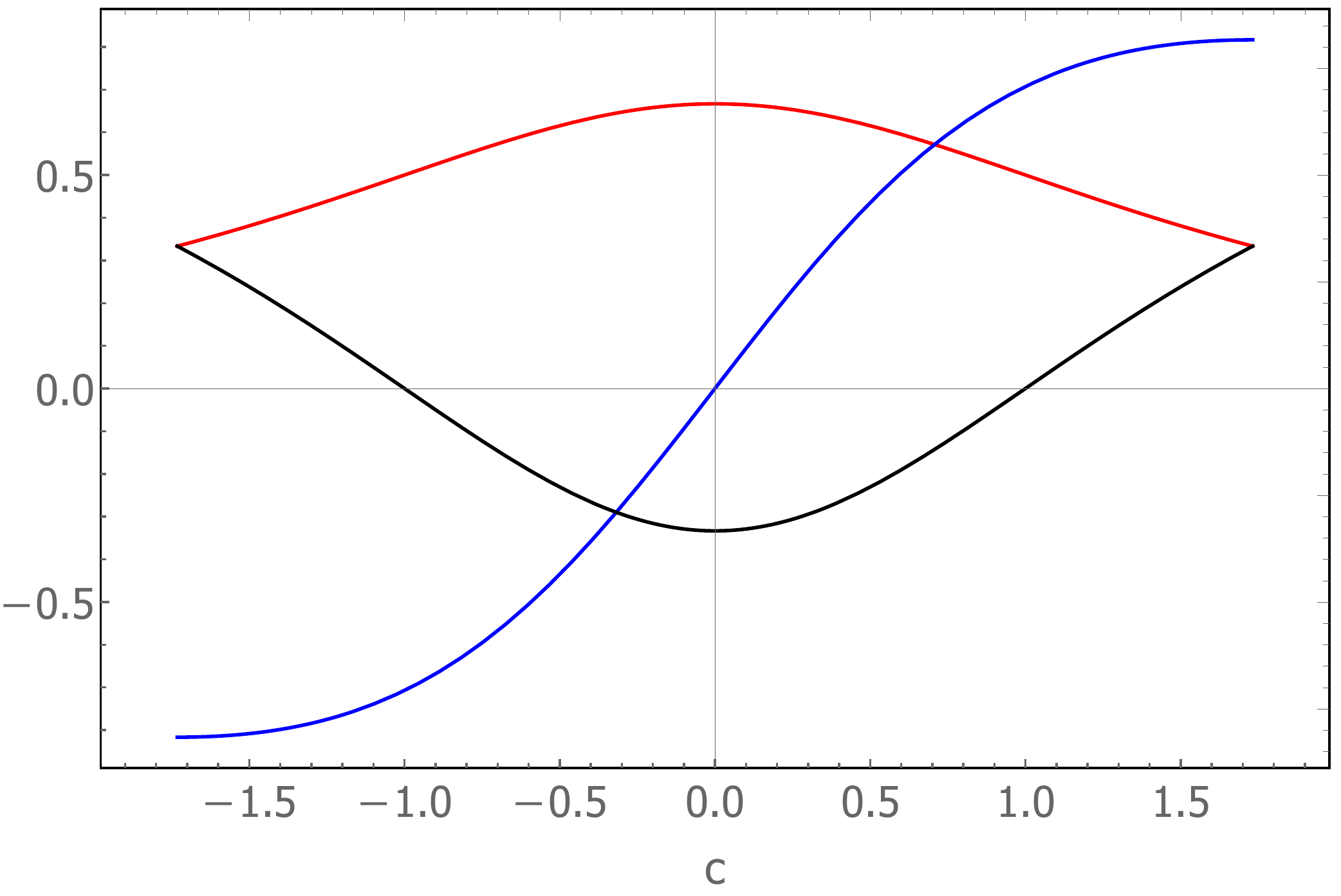}
\end{center}
\vspace{-0.6cm}
\caption{\small The Kasner exponents $p_t$ (black line), $p_x$ (red line) and $p_{\phi}$ (blue line) as functions of parameter $c$ in \eqref{eq:rel2}.
} 
\label{fig:prange}
\end{figure}

Another utility of this conserved charge comes from the fact that it gives a constraint relations between the parameters at the boundary, horizon and the singularity. Explicitly evaluating \eqref{conserved-charge} at the boundary, horizon and near the singularity gives the relation
\begin{eqnarray}
\label{eq:ccre}
	3 m_T - 4 \alpha \beta + \alpha^3 \lambda_3  = -\frac{4\pi T e^{-\chi_h/2}}{r_h^2} = f_1 e^{-\chi_1/2}\, (c^2-3)\,,
\end{eqnarray}
where we have used \eqref{eq:rel2}. 
Therefore we have $c^2 \leq 3$ which gives a bound of Kasner exponents
\begin{align}
\frac{1}{3} \leq p_x \leq \frac{2}{3}\,, ~~ -\frac{1}{3} \leq p_t \leq \frac{1}{3}\,,~~ -\sqrt{\frac{2}{3}} \leq p_\phi \leq \sqrt{\frac{2}{3}}\,.
\end{align}
Note that the singularity in Schwarzschild black hole saturates the upper bound of $p_x$ and the lower bound of $p_t$. The plot of Kasner exponents as a function of $c$ is shown in Fig.\ref{fig:prange}. Although  from the above relations, both $p_x$ and $p_t$ could be larger than zero due to the existence of the scalar field, it is interesting to note that for the parameters we considered we only find the case with negative $p_t$.

\subsection{Thermodynamics of black holes}

With the numerical solution constructed in the above subsection, we can compute the free energy density of the dual system. When there are multiple gravitational solutions, the system might experience phase transitions. In this subsection, we compute the thermodynamical quantities in the system under study. 

The thermodynamics of the Einstein-Scalar system is encoded in the appropriate renormalized Euclidean action. In our setup, the action must also take into account the more generalized nature of the boundary conditions on the scalar field due to alternative quantization and the double trace deformation. The action takes the form \cite{Faulkner:2010gj, Papadimitriou:2007sj}
\begin{align}\label{LorRenAction}
\begin{split}
	S &= \int_{\mathcal M} d^4x\; \sqrt{-g}\left[R + 6 - g^{ab}\pd_a\phi \pd_b \phi - V(\phi)\right]+ \int_{\partial \mathcal M} d^3x\; \sqrt{-\gamma}\,\left(2 K\right)  \\   &\hspace{1cm}+ \int_{\partial \mathcal M} d^3x\; \sqrt{-\gamma} \left(-4+ 2\phi\, n^a \pd_a \phi+(1+\kappa r_c)\phi^2 + \sigma\phi^3 -2 \lambda_3 \phi^3 \log r_c \right) 
	\end{split}
\end{align}
where we have set $16\pi G =1$, the boundary $\pd \mathcal{M}$ is at $r = r_c$ with $r_c\to 0$, $\gamma_{\mu\nu}$ and $ K$ are the induced metric and exterior curvature of boundary respectively. The parameter $\sigma$ is fixed by the fact that the cubic interaction term should not change the double trace boundary conditions as shown in appendix \ref{app:holoren}, and we have $\sigma =- 2\lambda_3$.

Using the ansatz \eqref{field-ansatz} and the near-boundary expansion \eqref{eq:nbexp}, it can be seen that the Euclidean on-shell action is given by
\begin{eqnarray}\label{On-shellEucAction}
	S^{E}_\text{on-shell} = \frac{V}{T} \left( m_T - \alpha^2 \kappa \right)
\end{eqnarray}
where $V$ is spatial volume. The density of free energy is therefore given by
\begin{eqnarray}\label{free-energy}
	f_{\kappa} \equiv \frac{F}{V} = \frac{TS^{E}_\text{on-shell}}{V}= m_T - \alpha^2  \kappa \,.
\end{eqnarray}
Note that $m_T$ is the coefficient in the metric field near AdS$_4$ boundary,  $\alpha$ is the VEV of the dual operator and $\kappa$ is the double trace deformation parameter. Therefore, with numerical solution obtained above, one can get the corresponding free energy density. For example, for the AdS planar black hole solution, it can be verified that $f_\text{Sch} = -\frac{1}{ r^3_h} = -\left(\frac{4\pi}{3}\right)^3 T^3$.


 It can also be checked that the free energy $\eqref{free-energy}$ also satisfies the thermodynamic relation
\begin{eqnarray}\label{fmT-thermo-relation}
	f_\kappa = \epsilon - T s
\end{eqnarray}
where $\epsilon$ and $s$ are the energy and entropy densities. The expression for the energy density can be read off from the Brown-York stress tensor at the boundary
\be
T_{\mu\nu}=2(K_{\mu\nu}-\gamma_{\mu\nu}K)+
\frac{2}{\sqrt{-\gamma}}\frac{\delta S_\text{c.t.}}{\delta\gamma^{\mu\nu}}
\ee
where $S_\text{c.t.}$ is the last term in (\ref{LorRenAction}). 
We obtain the energy density
\begin{eqnarray}
	\epsilon = \lim_{r_c\to 0}\sqrt{-\gamma} T^0_0=  -2m_T + 4 \alpha\beta  - \sigma \alpha^3 - 3\lambda_3 \alpha^3 -  \alpha^2 \kappa\,. 
\end{eqnarray}
The expression for $Ts$ can be obtained in terms of the boundary parameters using the relation (\ref{eq:ccre}) from the conserved charge, we have
\begin{eqnarray}
	-Ts = 3m_T - 4\alpha\beta + \lambda_3 \alpha^3\,.
\end{eqnarray}
This concludes the proof of the thermodynamic relation \eqref{fmT-thermo-relation}.

\section{Behavior of the singularity across the phase transition}
\label{sec:pt}

In the previous section we have collected all the necessary ingredients to solve the system, we shall show the numerical results of the system in this section. We will first show the example with second order phase transition and analyze the behavior of the singularity across the transition. Then we will study the first order phase transitions and also analyze the singularity of the black hole solutions. The probes of the black hole singularities in the context of AdS/CFT will also be studied.

\subsection{Second order phase transition}
\label{ss:2nd}

It has been shown in \cite{Faulkner:2010gj,Mefford:2014gia} that when the potential $V(\phi)$ in \eqref{EinScalarAction} is symmetric under $\phi\to-\phi$, i.e. $\lambda_3=0$, there is a second order phase transition between Schwarzschild black hole and hairy black hole for $\kappa < 0$. Here we generalize these results with a different choice of the potential for the scalar field and focus on the behaviors of the black hole singularities. 

\begin{figure}[h!]
\begin{center}
\includegraphics[
width=0.475\textwidth]{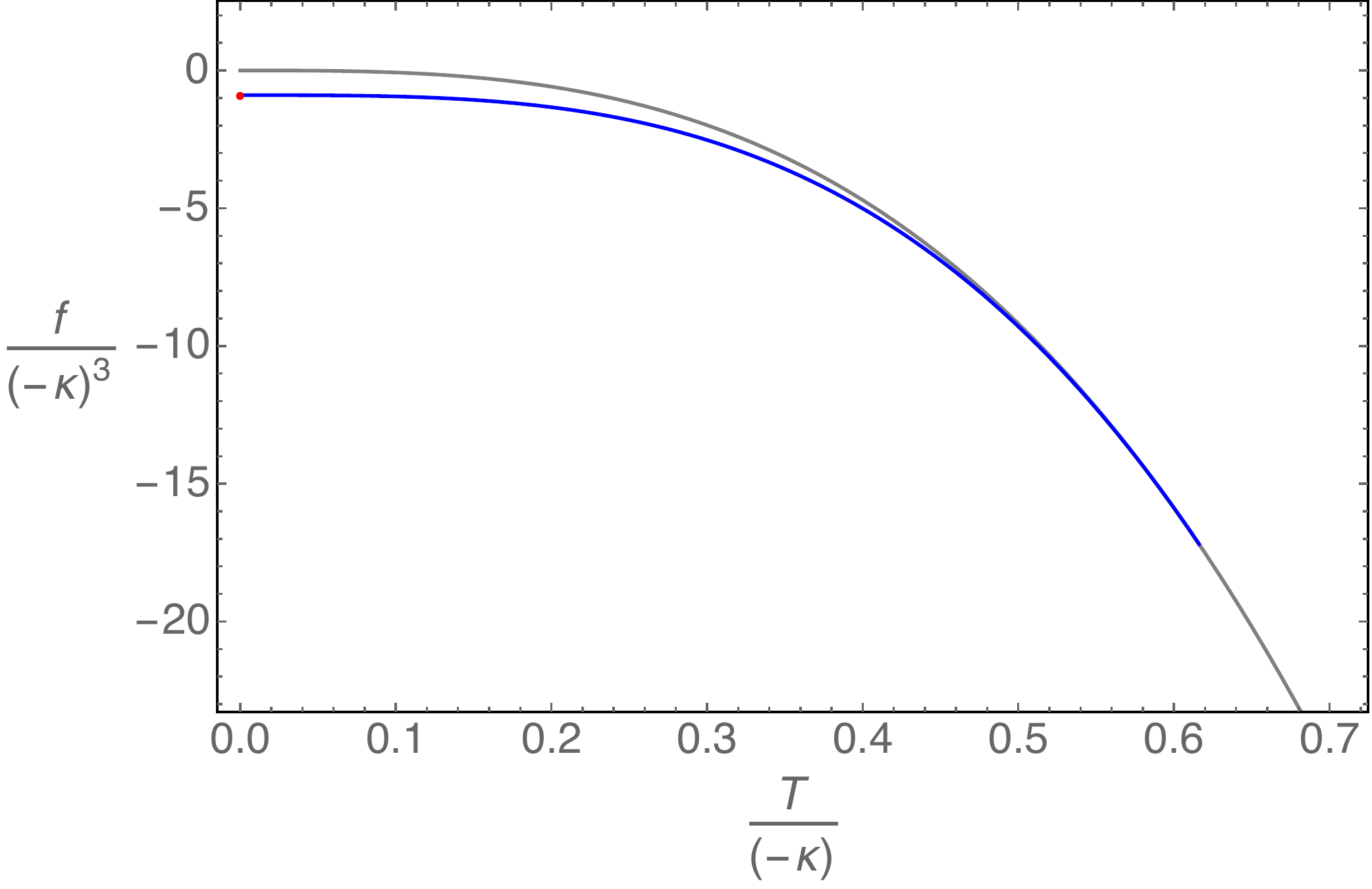}\,~~~
\includegraphics[
width=0.47\textwidth]{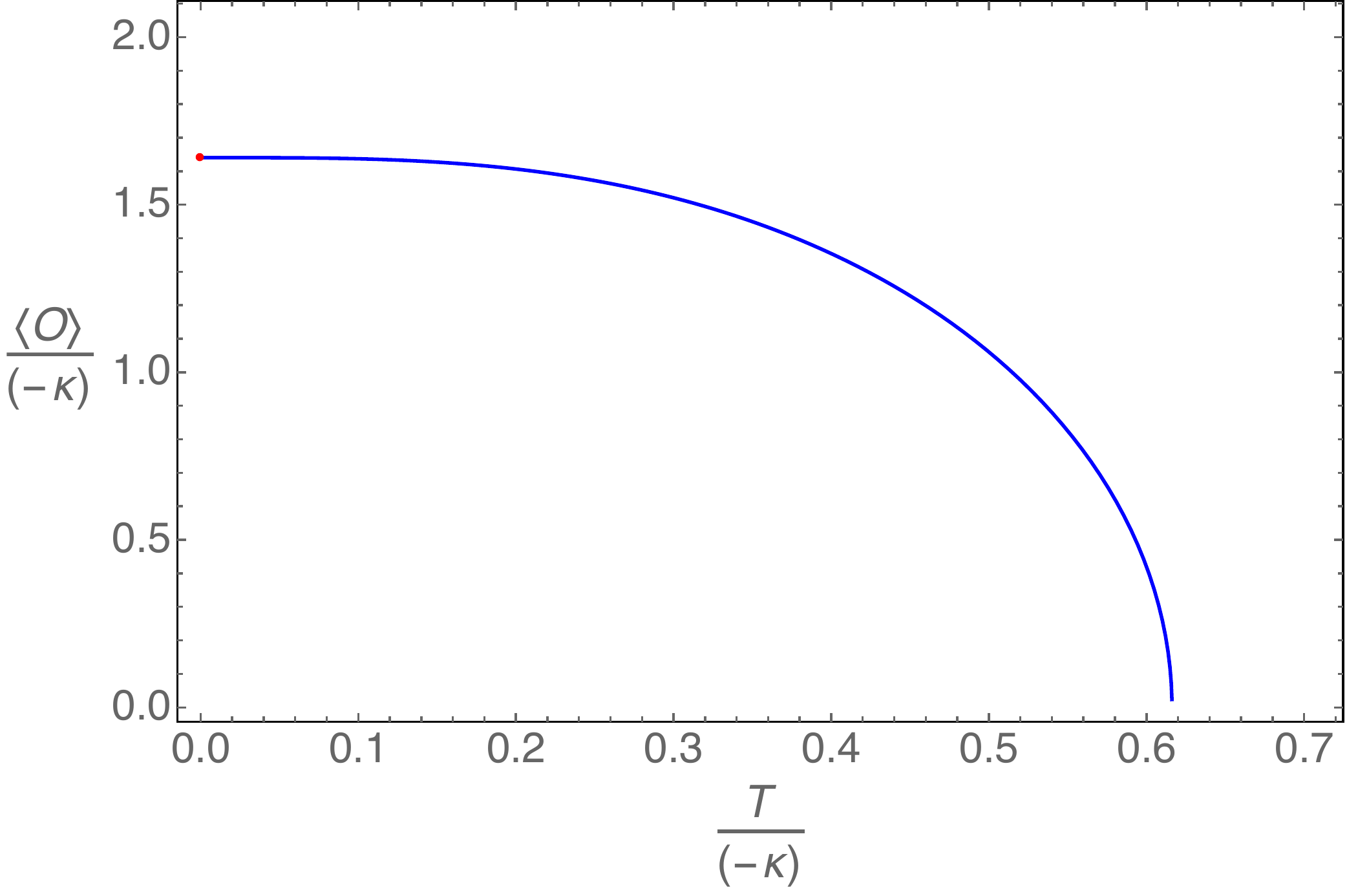}
\end{center}
\vspace{-0.7cm}
\caption{\small {\em Left:} The free energy across second order phase transition. The grey and blue lines are the free energy densities of Schwarzschild black hole and hairy black hole respectively. The red dot is the free energy of the hairy black hole at zero temperature. {\em Right:} The order parameter $\langle O \rangle/(-\kappa)$ as a function of $T/(-\kappa)$.} 
\label{fig:vev-2nd}
\end{figure}

We choose parameters $\lambda_3=0, \lambda_4=1/10$ in the scalar potential (\ref{eq:potential}) and solve the system with double trace deformation. With the boundary condition that the source  $\kappa\alpha-\beta$ should vanish, the hairy black hole solution can only be found below a critical $T_c/(-\kappa)\approx 0.616$ which takes almost the same values as in \cite{Mefford:2014gia}. This is due to the fact that close to the phase transition, the value of the scalar field is very small and the temperature can be analytically computed at the leading order in the expansion of the scalar field. The free energy densities of the black hole solutions and the expectation values of the scalar operator as a function of $T/(-\kappa)$ are shown is Fig. \ref{fig:vev-2nd}. In the left plot of Fig. \ref{fig:vev-2nd}, the blue line is the free energy of the hairy black hole while the grey line is the free energy of the Schwarzschild black hole. It is seen that below $T_c/(-\kappa)$ the hairy black hole is more stable. The right plot of Fig. \ref{fig:vev-2nd} shows the behavior of the condensation, i.e. the order parameter of the phase transition, as a function of $T_c/(-\kappa)$ in the hairy black hole. When we lower the temperature, the condensate increases up to a constant value at zero temperature.  Close to the transition temperature $T\to T_c$, we have $\langle{O}\rangle/(-\kappa)\propto (1-T/T_c)^{1/2}$ and $\delta f/(-\kappa)^3=(f_\text{Sch}-f_\kappa)/(-\kappa)^3 \propto (1-T/T_c)^{2}$. From these behaviors, we conclude that the phase transition is of second order. 

\begin{figure}[h!]
\begin{center}
\includegraphics[height=6cm, width=0.63\textwidth]{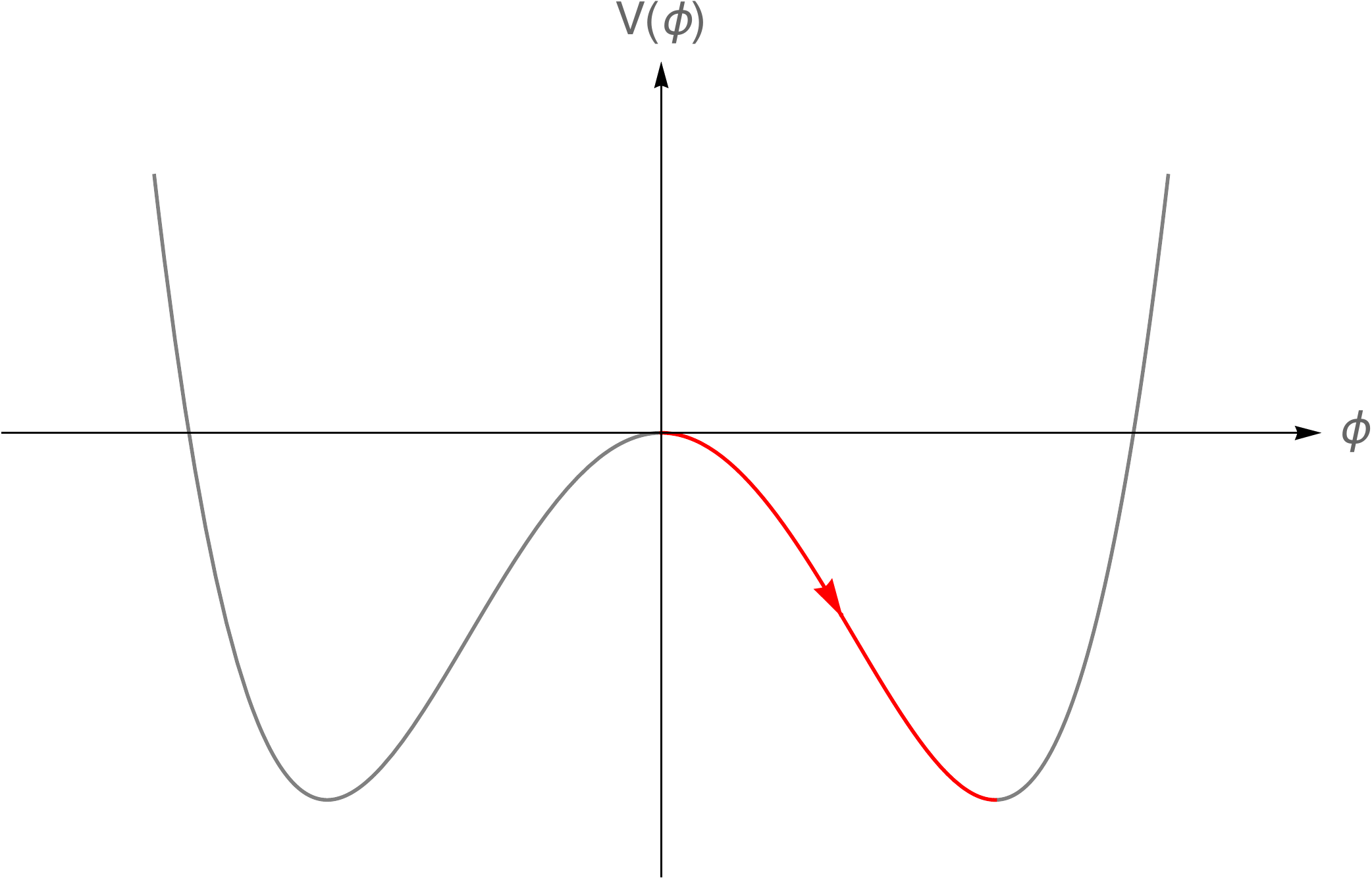}
\end{center}
\vspace{-0.6cm}
\caption{\small When the temperature is lowered down, the flow of the near horizon value of the scalar field is shown in the red line with arrow.} 
\label{fig:phih-2nd}
\end{figure}

At zero temperature, the strategy for constructing the bulk solution should be modified because now the near horizon conditions are different from \eqref{eq:nhexpansion}. In this case, the 
near horizon expansion as $r\to\infty$ are 
\begin{align}
	\begin{split}
	f &= 1+ \frac{1}{6\lambda_4}+ \frac{a_1^2+96\lambda_4 + 6\lambda_4 a_1^2}{48\lambda_4(a_1-3)}\, \phi_0^2\,r^{a_1}\,,\\
	\chi &= \frac{a_1}{4}\,\phi_0^2\,r^{a_1}\,,\\
	\phi &= \frac{1}{\sqrt{\lambda_4}} +  \phi_0\,r^{\frac{a_1}{2}}\,,
		\end{split}
\end{align}
where $
	a_1 = \frac{3+18\lambda_4 -  \sqrt{9+204\lambda_4 +900 \lambda_4^2}}{1+6\lambda_4}$
which is $3-\sqrt{15}$ for the choice of $\lambda_4=1/10$. At the leading order the above near horizon geometry is AdS$_4$, and the irrelevant deformation flow the geometry to AdS$_4$. 
Note that from the scaling symmetry (\ref{scalsym2}), $\phi_0$ can be rescaled to arbitrary value by scaling the radial coordinate $r$. Therefore only the sign of $\phi_0$ is crucial to flow the near horizon to AdS$_4$ boundary and we have a unique solution at zero temperature, which is an AdS$_4$ to AdS$_4$ domain wall with different AdS radius.  
The free energy of the zero temperature solution is shown as a red dot in the left plot of Fig. \ref{fig:vev-2nd}. It can be seen that the hairy black hole is always more stable than Schwarzschild black hole. The trajectory of the near horizon value of the scalar field $\phi$ when the temperature is lowered down can be seen from Fig. \ref{fig:phih-2nd}. 
At the transition temperature $T_c/(-\kappa)$, we have $\phi_h=0$ and at zero temperature we have $\phi_h=1/\sqrt{\lambda_4}$. Note that there is a $Z_2$ symmetry $\phi\to-\phi$ for the system. We could have other branch of solution with the same behavior as above. 

\begin{figure}[h!]
\begin{center}
\includegraphics[
width=0.64\textwidth]{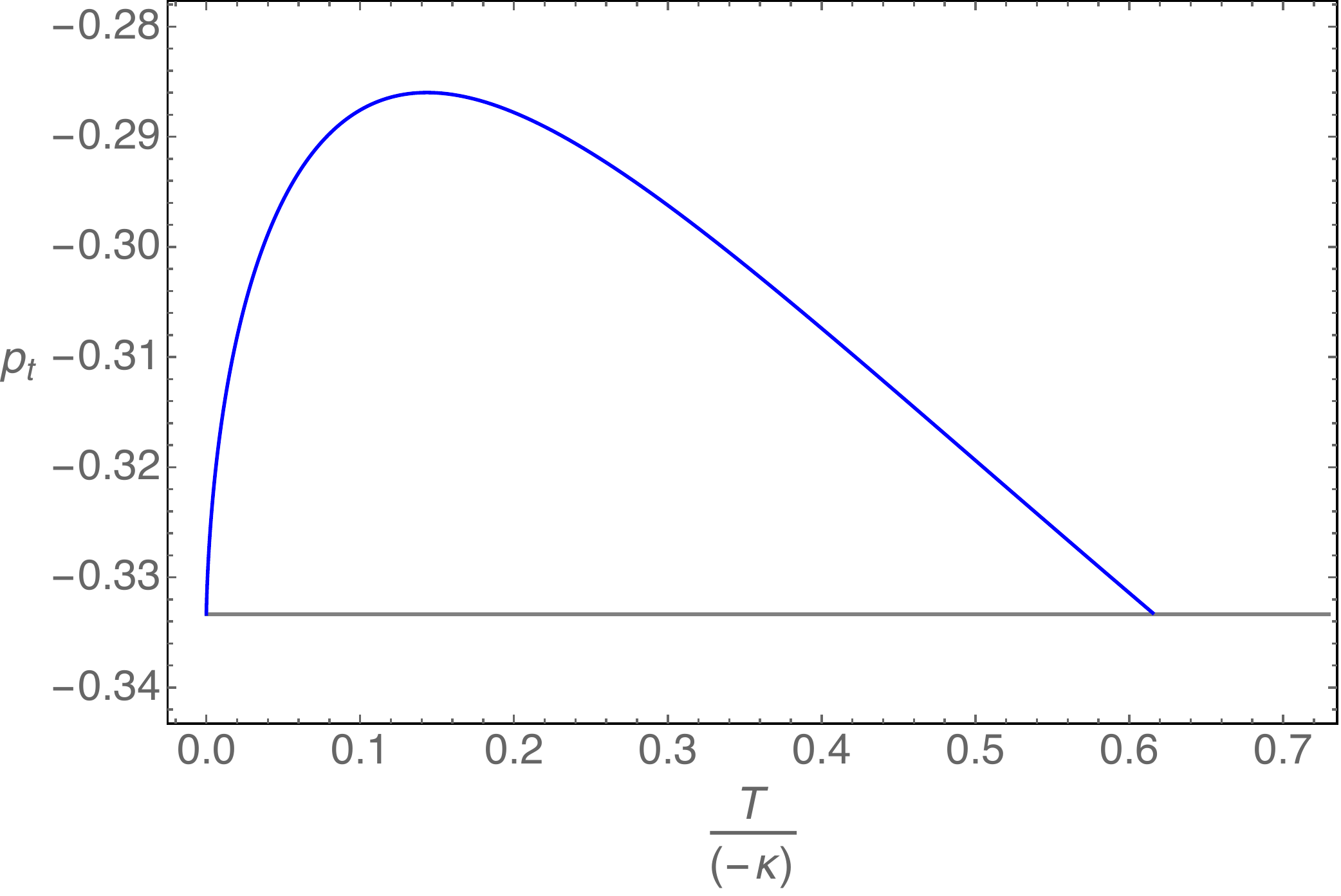}
\end{center}
\vspace{-0.7cm}
\caption{\small The Kasner exponent $p_t$ as a function of $T/(-\kappa)$ for $\lambda_3=0, \lambda_4=1/10$. The grey line is $p_t$ of Schwarzschild singularity which is $-1/3$. The blue line is for $p_t$ of the hairy black hole solution. When $T/(-\kappa)$ goes to zero, $p_t$ goes to $-1/3$
.} 
\label{fig:pt-2nd}
\end{figure}

Next, we study the Kasner exponent across the second order phase transition above. With the near horizon conditions for the hairy black hole solution, we can integrate the system towards the singularity from the horizon, and obtain the Kasner exponents numerically. One of the Kasner exponents, $p_t$ as a function of $T/(-\kappa)$,  is shown in Fig. \ref{fig:pt-2nd}, where the blue line is for hairy black hole while the grey line is for Schwarzschild black hole. One finds that during the second order phase transition at $T_c/(-\kappa)$, the Kasner exponent $p_t$ is continuous while the first derivative of $p_t$ with respect to $T/(-\kappa)$ is not continuous. Since from \eqref{eq:rel2} only one of the three Kasner exponents is independent, we conclude that the Kanser exponents are continuous and non-differentiable during the second order phase transitions.

The above behavior of the Kasner exponents across the second order mean field phase transition seems universal. Firstly, the same behavior can be found in other holographic examples. The Kasner exponents are also studied in the Einstein-Maxwell theory coupled to neutral scalar field in \cite{Hartnoll:2020rwq} and in Fig. 6 of \cite{Hartnoll:2020rwq}  one example of phase transition was shown. From this we could see that Kasner exponent is continuous while its first derivative with respect to $T/\mu$ is not for a continuous phase transition if we identify $p_t=1$ for the singularity of RN AdS black hole. Secondly, one naive analytical argument could follow from the fact that the metric and matter fields are analytical functions of radial coordinate $r$ at any temperature.\footnote{We thank Run-Qiu Yang for a
helpful discussion on this point.} Note that close to the transition temperature the scalar field is very small. From the side of the hairy black hole we have $\mathcal{O}\propto (1-T/T_c)^{1/2}$, thus we expect that $\phi(r)\propto (1-T/T_c)^{1/2}$. From \eqref{eq:rel1} we have 
$c\propto (1-T/T_c)^{1/2}$. Therefore, from \eqref{eq:rel2} we have $p_t+\frac{1}{3}\propto (1-T/T_c)$. This means that in general for a second order  mean field phase transition, the Kasner exponent is continuous while the first order derivative with respect to $T$ is not continuous.



\subsection{First order phase transition}
\label{ss:1st}

Now we consider the case with first order phase transition and study the behavior of the singularity across the phase transitions. We choose a non-symmetric scalar potential, and without loss of generality we set $\lambda_3=1/8, \lambda_4=1/10$ in  (\ref{eq:potential}). 

\begin{figure}[h!]
\begin{center}
\includegraphics[
width=0.475\textwidth]{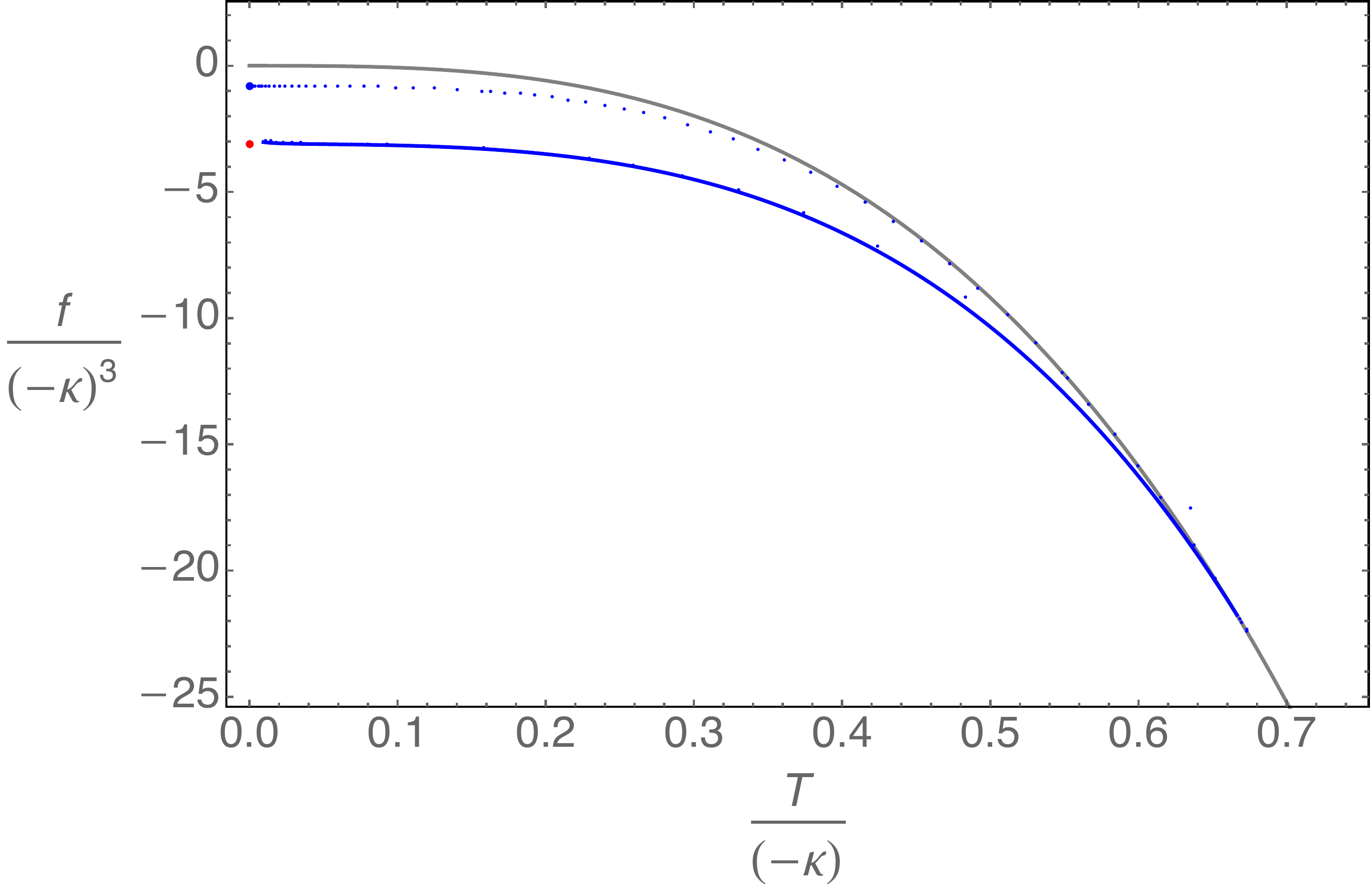}
\,~~~~
\includegraphics[
width=0.47\textwidth]{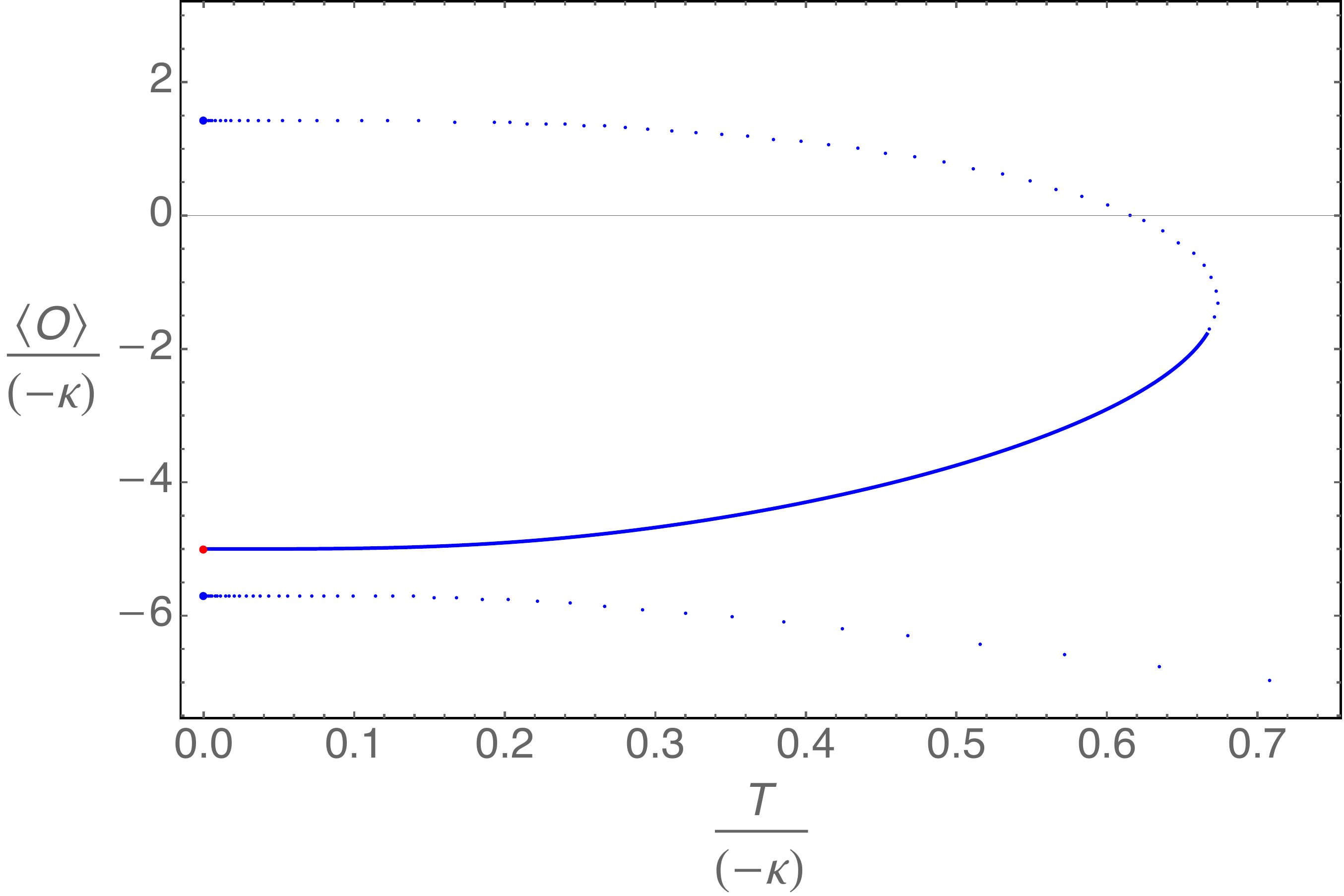}
\end{center}
\vspace{-0.6cm}
\caption{\small {\em Left:} The free energy of the black hole solutions with $\lambda_3=1/8$ and $\lambda_4=1/10$. The grey line is for Schwarzschild solution and the other lines are for hairy black hole solutions. {\em Right:} The VEV $\langle O \rangle$ as a function of $T/(-\kappa)$. In both plots, the solid blue line is for stable hairy black hole solution while the dotted blue lines are for unstable hairy black hole solutions.} 
\label{fig:vev-1st}
\end{figure}

We find that the hairy black hole solution can exist at arbitrary value of $T/(-\kappa)$. We plot the free energy of the hairy black hole solutions (in blue) and Schwarzschild solution (in grey) in the left plot of Fig. \ref{fig:vev-1st}. 
When we lower the temperature, we find that there is a first order phase 
transition $T_{c1}/(-\kappa) \approx 0.666$ where the first order derivative of free energy density with respect to $T/(-\kappa)$ is discontinuous. 
The behavior of the order parameter $\langle{O}\rangle/(-\kappa)$ as a function of $T/(-\kappa)$ is shown in the right plot of Fig. \ref{fig:vev-1st} where the solid lines are for stabler phases. 
At the transition temperature $T_{c1}/(-\kappa)$, the condensation $\langle{O}\rangle/(-\kappa)$ has a jump to a negative value and then decreasing until a constant value when we lower temperature. Therefore the  system goes through a first order phase transition. 

\begin{figure}[h!]
\begin{center}
\includegraphics[height=6cm, width=0.65\textwidth]{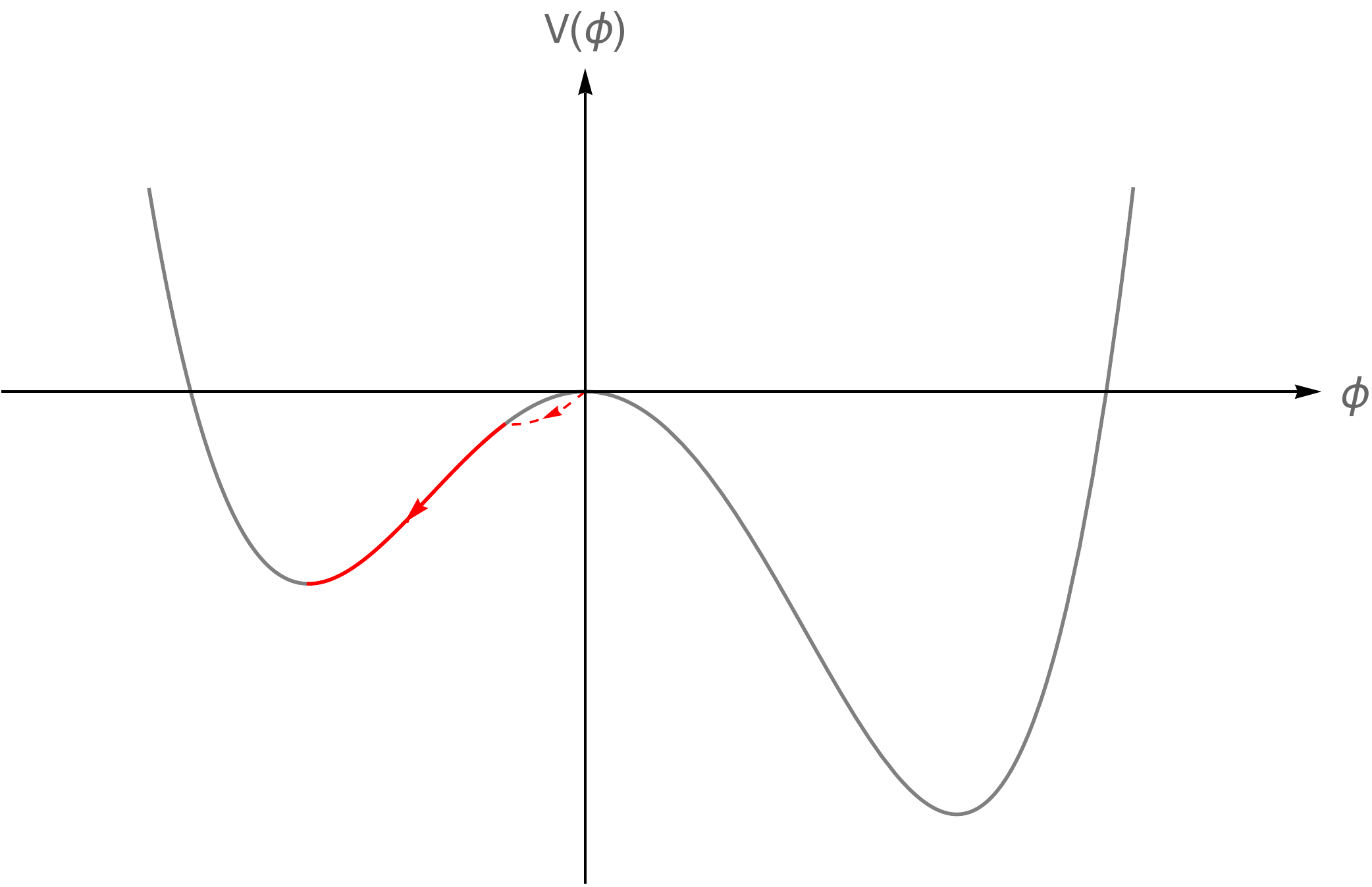}
\end{center}
\vspace{-0.6cm}
\caption{\small The red solid line with arrows denote the trajectory of the near horizon value of the scalar field when we lower the temperature.} 
\label{fig:phih-1st}
\end{figure}

At zero temperature, we have the near horizon boundary condition as $r\to\infty$ 
\begin{align}
\label{eq:nhasT=0}
	\begin{split}
	f &= c_0 
	+ c_1 \, \phi_0^2\,r^{a_2}\,,\\
	\chi &= c_2 \, \phi_0^2\,r^{a_2}\,,\\
	\phi &= \frac{3\lambda_3-\sqrt{9\lambda_3^2+64\lambda_4}}{8\lambda_4} +  \phi_0\,r^{\frac{a_2}{2}}\,,
	\end{split}
\end{align}
where $c_0$, $c_1, c_2$ and $a_2$ are functions of $\lambda_3$ and $\lambda_4$. For the case we considered, i.e. $\lambda_3=1/8$ and $\lambda_4=1/10$, we have $(c_0, c_1,c_2,a_2) = (2.135, -0.503,-0.235,-0.942)$. Note that $\phi_0$ can be rescaled to arbitrary value by the symmetry (\ref{scalsym2}). We choose specific $\phi_0$ to fix $\kappa = -1$.
The free energy at zero temperature from the above solution $f_{\kappa 0}/(-\kappa)^3= -3.125$ is smaller than the free energy of Schwarzschild solution. 
Note that for boundary condition (\ref{eq:nhasT=0}), there are two $\phi_0'$s corresponding to $\kappa = -1$. We have chosen the more stabler  one and the free energy of the other one is $f_{\kappa}/(-\kappa)^3= -3.114$.
Note that in (\ref{eq:nhasT=0}) we chose the near horizon value of $\phi$ to be the local minimal of the potential of the scalar field. Since 
there exist two minima of the potential, there are two different near horizon conditions and the other one has the form of $\phi=\frac{3\lambda_3+\sqrt{9\lambda_3^2+64\lambda_4}}{8\lambda_4} +  \tilde{\phi}_0\,r^{\frac{\tilde{a}_2}{2}}$. However, the free energy of this one is $ f_{\kappa}/(-\kappa)^3 = -0.803$ and it is bigger than $f_{\kappa 0}/(-\kappa)^3$. 
The trajectory of the near horizon value of the scalar filed when we lower temperature is shown as the solid red line with arrow in Fig. \ref{fig:phih-1st}. Different from the previous symmetric case, now the potential is not symmetric and the trajectory of the near horizon value of the scalar filed is unique. When we lower down the temperature, at $T_{c1}/(-\kappa)$, the near horizon value $\phi_h$ jumps from 0 to a specific negative number and then flows to the left minimal of the potential witch corresponding to the truly zero temperature solution.
The stabler phase can be seen from the horizon behavior of the scalar field.  
It is interesting to compare with the study of holographic phase transitions in \cite{Gursoy:2018umf,Bea:2018whf} without double trace deformation, where a jump of the near horizon value of scalar field was also found and is related to the existence of the noncontinuous phase transition. 

\begin{figure}[h!]
\begin{center}
\includegraphics[
width=0.46\textwidth]{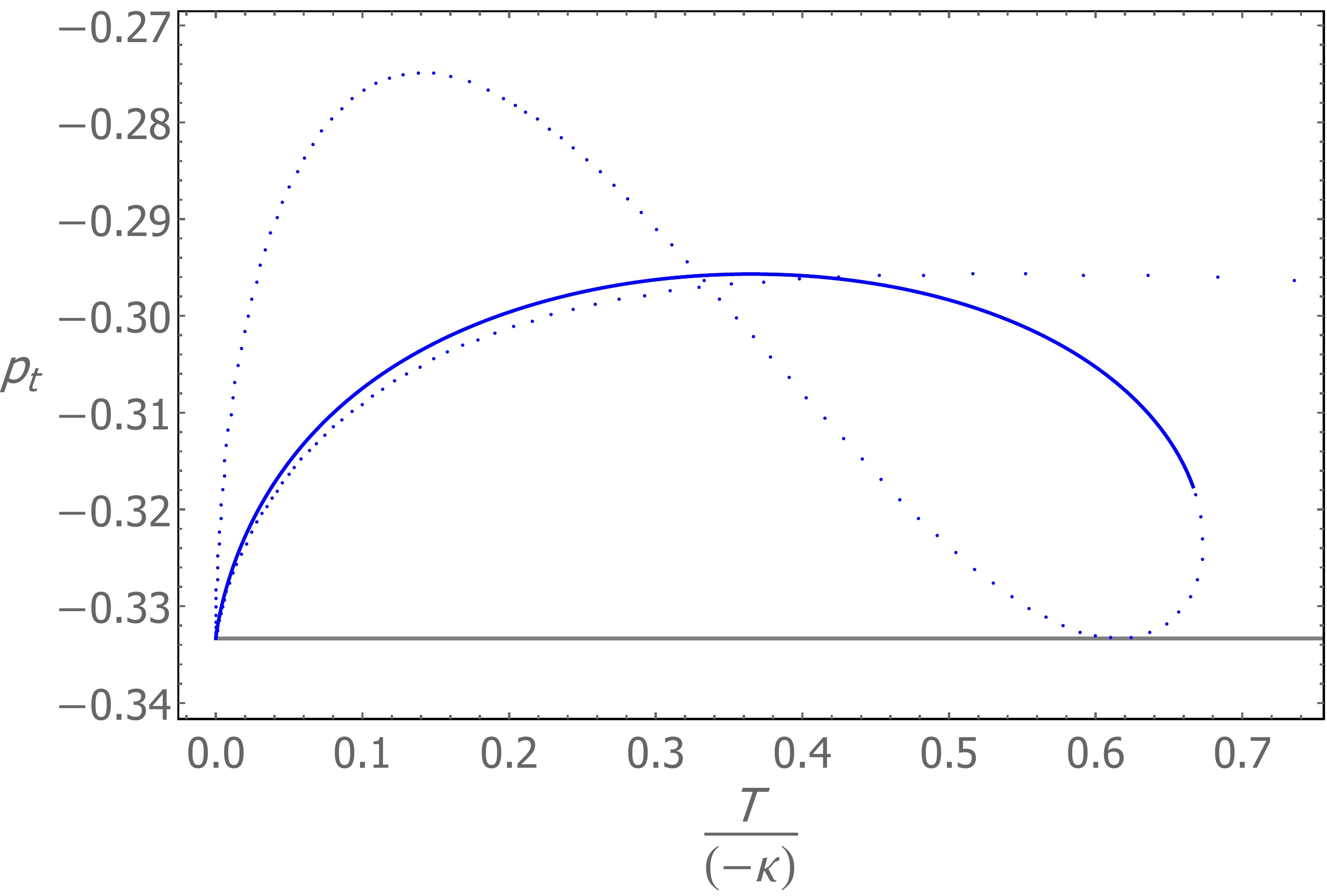}\,~~~
\includegraphics[
width=0.46\textwidth]{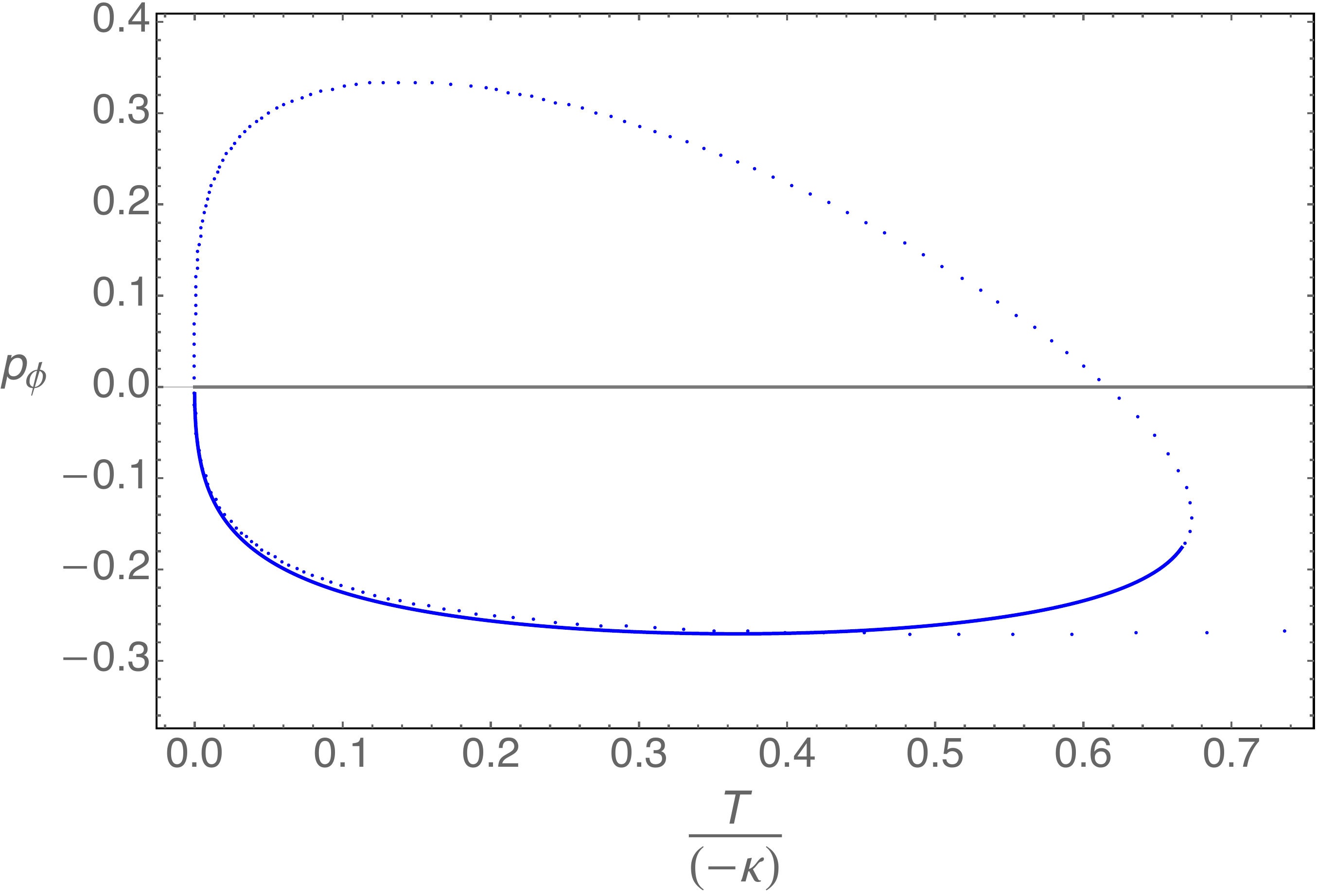}
\end{center}
\vspace{-0.7cm}
\caption{\small The Kasner exponents $p_t$ ({\em left}) and $p_\phi$ ({\em right}) of the black hole solutions as a function of $T/(-\kappa)$. In both plots, the gray line is for Schwartzschild black hole, the solid blue line is for stable hairy black hole solution while the dotted blue lines are for unstable hairy black hole solutions.} 
\label{fig:pt-1st}
\end{figure}

The Kasner exponents $p_t$ and $p_\phi$ of the singularities are shown in Fig. \ref{fig:pt-1st} where the Kasner exponents of the hairy black hole and Schwarzschild black hole are in blue and grey respectively. Below $T_{c1}/(-\kappa)$, the Kasner exponents of the stable phases are in solid blue. Different from the case of continuous phase transition, when the phase transition is of first order, the Kasner exponents are discontinuous.  

It is natural to expect that this discontinuity of the Kasner exponent is a universal behavior for discontinuous phase transitions. Firstly, the exactly same behavior can be found in other example of first order Hawking-Page transition. 
For holographic CFTs on $S^{d-1}$ at finite temperature, there is a well-known first order Hawking-Page transition \cite{Witten:1998zw} between the theory whose dual geometry is described by a black hole in global AdS$_{d+1}$ at high temperature  
and  the theory which is dual to a thermal AdS spacetime at low temperature. 
The Schwarzschild black hole in AdS has Kasner exponents $(p_t, p_{\omega_i})=(\frac{2}{d}-1,\frac{2}{d})$ with $i=1,\dots, d-1$ and $\omega_i$ the $i$-th spherical coordinate. It is easy to check that the Kasner relations $p_t+(d-1)p_{\omega_i}=p_t^2+(d-1)p_{\omega_i}^2=1$ are satisfied. In the thermal AdS we do not have any singularity and the geometry is no longer of Kasner form (\ref{eq:kasner}). In this sense the Kasner exponents of the singularity are discontinuous during the first order phase transition. From the holographic first order phase transition between the AdS Schwartzschild black hole in planar coordinate and AdS-soliton one can obtain the same conclusion. Secondly, in the Einstein-scalar theory with double trace deformation, during the first order phase transition, the scalar field is discontinuous, therefore we expect that the value $c$ in the near singularity behavior of the scalar field  \eqref{eq:rel1} is discontinuous. From \eqref{eq:rel2} we know that the Kasner exponents are discontinuous during the first order phase transitions.

\subsection{Probes of the black hole singularity via geodesics}
\label{ssec:probe}

In this subsection, we will study some observable quantities of boundary field theory from which we can read the information about Kasner exponents of the black hole's singularity.  We shall focus on the geodesics since the bulk geodesics can approach the singularity of the black hole in certain limit. In the context of AdS/CFT duality, the length of bulk spacelike geodesics corresponds to the correlation functions of the operators in the large conformal dimensional limit of dual boundary field theory \cite{Fidkowski:2003nf,Festuccia:2005pi}. Meanwhile, the connection between the proper time from the black hole horizon to the singularity and one-point functions were proposed in \cite{Grinberg:2020fdj}. In the following we will compute these two types of geodesics respectively. 

 We consider the radical geodesics for which $g_{tt} \dot{t}^2 + g_{rr} \dot{r}^2 = \varepsilon$, where the dot denotes the derivative with respect to the proper time $\tau$ and $\varepsilon =-1,0,1$ gives timelike, null and spacelike geodesics respectively. There is a conserved charge $E = -g_{tt} \dot{t}$ along the geodesic.  For timelike or null geodesics $E$ characterize the conserved energy of the particle.
The equation of motion of the geodesic is
\begin{eqnarray}
	\frac{E^2}{g_{tt}}+g_{rr} \dot{r}^2 = \varepsilon
\end{eqnarray}
from which we obtain
\begin{align}
\label{eq:geo}
 \frac{dr}{d\tau}=\sqrt{\frac{\varepsilon g_{tt} - E^2}{g_{tt} g_{rr}}}\,.
\end{align}
In the following we will solve this equation for timelike geodesics and spacelike geodesics respectively.

\subsubsection{Radial timelike geodesics}
Recently, it was proposed in \cite{Grinberg:2020fdj} that the proper time $\tau_s$ from the black hole horizon to the singularity is related to the expectation value of certain operators in the large mass limit via $\langle \mathcal{O} \rangle\propto e^{-i m\tau_s-ml_\text{hor}}$, 
where $m$ is a complixified mass with $\text{Im}(m)<0$ and $l_\text{hor}$ is the distance from horizon to the boundary. Therefore, from the one point function of the operator one can extract the information of Kasner exponent once we know the relation between $\tau_s$ and the Kasner exponents. 

\begin{figure}[h!]
\includegraphics[
width=0.46\textwidth]{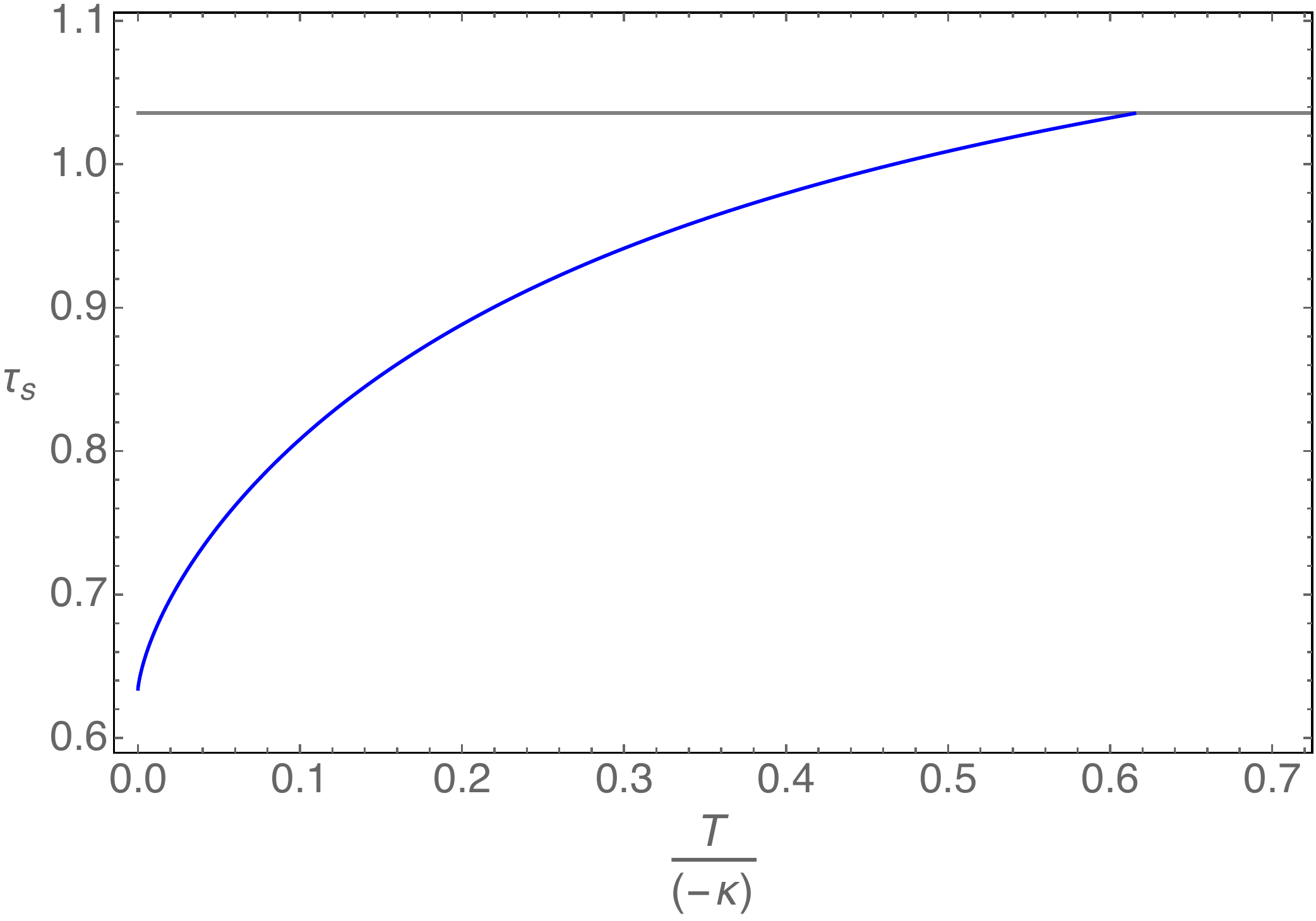}~~
\includegraphics[
width=0.47\textwidth]{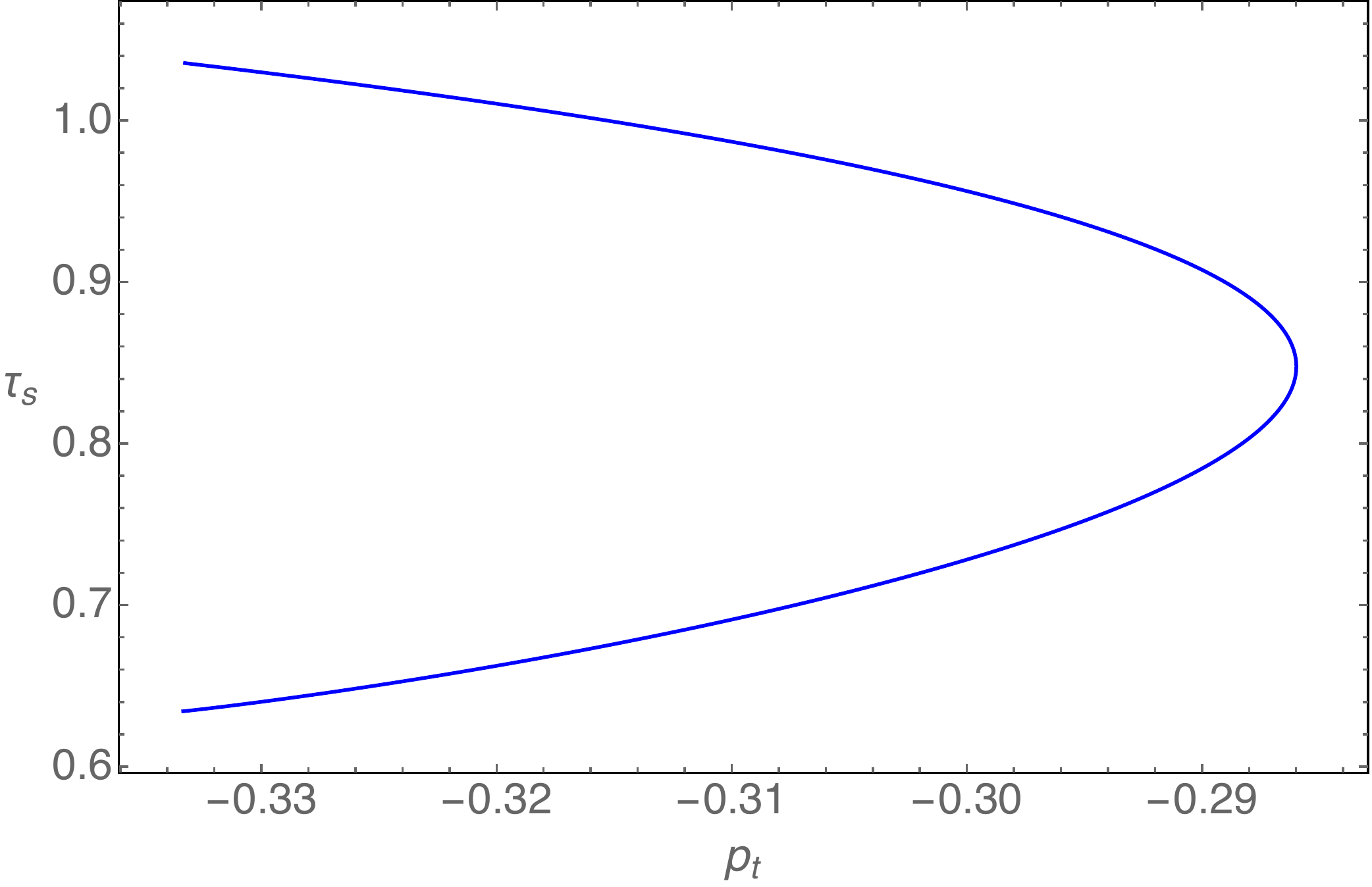}\\
\vspace{0.1cm}
\includegraphics[
width=0.46\textwidth]{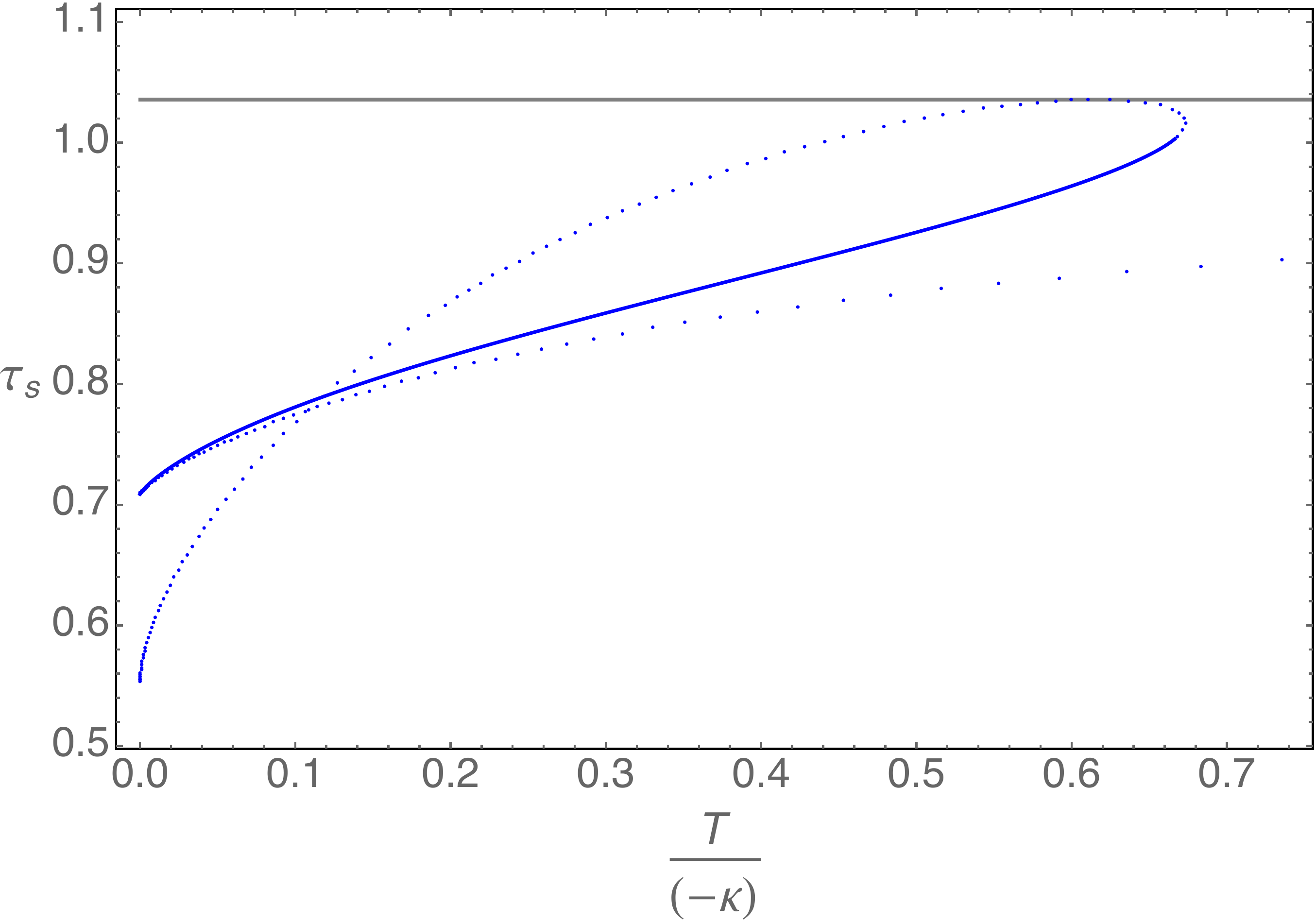}~~
\includegraphics[
width=0.47\textwidth]{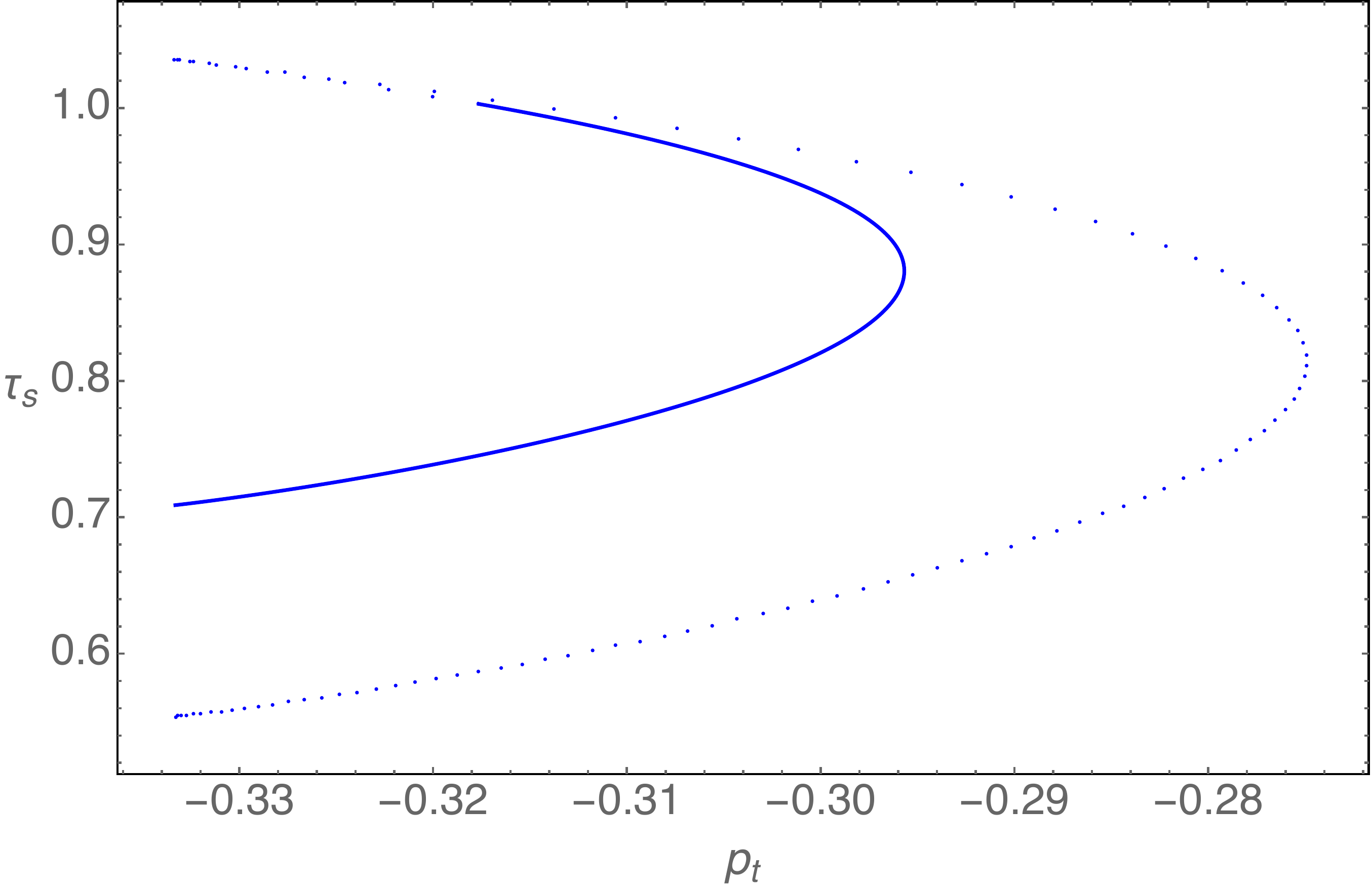}
\vspace{-0.2cm}
\caption{\small The proper time $\tau_s$ as a function of $T/(-\kappa)$ ({\em  left}) and $p_t$ ({\em right}) for the model in subsection \ref{ss:2nd} ({\em top plots}) and in subsection \ref{ss:1st} ({\em down plots}). For these plots, the grey line is for Schwartzschild black hole, the solid blue line is for stable hairy black hole solution while the dotted blue lines are for unstable hairy black hole solutions.} 
\label{fig:taus-2nd}
\end{figure}

For timelike geodesics we have $\varepsilon = -1$ in (\ref{eq:geo}). Plugging in the ansatz for the metric we obtain 
\begin{eqnarray}
	\frac{d\tau}{dr} = \frac{e^{-\chi/2}}{r^2 \sqrt{-E^2-(e^{-\chi} f/r^2)}}\,.
\end{eqnarray}
The proper time from horizon to singularity of a particle with $E=0$ is
\begin{eqnarray}
	\tau_s = \int_{r_h}^{\infty}d\tau  = \int_{r_h}^{\infty} \frac{dr}{r \sqrt{ - f(r)}}\,.
\end{eqnarray}
The proper time $\tau_s$ as a function of $T/(-\kappa)$ and $p_t$ are shown in Fig. \ref{fig:taus-2nd}. The left two  plots characterize that when we lower the temperature, the behavior of the $\tau_s$, or equivalently the expectation value of the operator $\mathcal{O}$. For this $\tau_s$, one can extract the information of the Kasner exponent $p_t$ according to the right plots in Fig. \ref{fig:taus-2nd}.

\subsubsection{Radial spacelike geodesics}

The spacelike geodesics are well studied in the literature as they are the relevant geodesics for the computation of correlators in AdS/CFT \cite{Fidkowski:2003nf,Festuccia:2005pi}.  A significant difference between the spacelike geodesics from the timelike geodesic is that they do not end in singularity. Instead the geodesics reach a finite radius $r_*$, namely the turning point, which is controlled by the energy $E$, after which it bounces back and goes to the other side of the eternal black hole. Nevertheless, when $E\to\infty$, the turning point will approach very close to the singularity and therefore in this limit one could extract the information of the singularity with the spacelike geodesics. 

For the spacelike geodesics, we set $\varepsilon =1$ in (\ref{eq:geo}). The regularized proper length is given by
\begin{eqnarray}
	L = 2 \int_{r_c}^{r_*} dr \frac{e^{-\chi/2}}{r^2 \sqrt{E^2 + \frac{1}{r^2}fe^{-\chi}}} + 2 \log r_c
\end{eqnarray}
where $r_*$ is the turning point which is given by $g_{tt}(r_*) = E^2$ and for real $E$, $r_*$ lies inside the horizon, and $r_c$ is the cutoff close to the AdS boundary. As $E$ becomes large, $r_*$ tends towards the singularity. At large $E$, most of the contribution comes from the boundary and we may approximate the integral by an expansion in $1/E$.\footnote{One subtlety of the expansion is that the limit $r_c\to 0$ and $E\to\infty$ do not commute with each other (see \cite{Frenkel:2020ysx} for details).} The resulting expansion is of the form
\begin{align}
	\begin{split}
		L &= 2 \log \frac{2}{E} + \frac{l_1}{E} + \frac{\alpha^2}{2}\frac{1}{E^2} + \frac{l_3}{E^3} + \Big(m_T - 2\alpha\beta + \alpha^3\lambda_3 -  \alpha^3 \lambda_3 \log 2 \Big)\,\frac{\log E}{E^3} \\ &\hspace{4cm}- \frac{1}{2}\alpha^3 \lambda_3\, \frac{\log^2 E}{E^3} + l' E^{\frac{1}{p_t}}\,,
	\end{split}
\end{align}
where the coefficients $l_1,l_3$  depend on the metric fields along the radial direction and
\begin{align}
	\begin{split}
		l' &= \sqrt{\pi} (p_t-1) \frac{e^{\frac{\chi_1}{2p_t}}}{f_1^{\frac{p_t+1}{2p_t}}} \frac{\Gamma(\frac{p_t+1}{2p_t})}{\Gamma(\frac{1}{2p_t})}\,,
	\end{split}
\end{align}
where $\chi_1$ and $f_1$ are defined in \eqref{eq:rel1}. The information of the Kasner exponent of singularity is encoded in the non-analytic part of the proper length $L$. 

From the AdS/CFT correspondence, the proper length of the spacelike geodesic is related to the correlators of the operators in the large conformal dimension limit. Therefore, in principle, from the correlation functions of the field theory one could obtain the Kasner exponent $p_t$.

\section{Conclusion and discussion}
\label{sec:condis}
We have studied the behaviors of the singularities in Einstein-scalar theory with a double trace deformation in which the second order and the first order phase transitions could be realized. We found that when the phase transition is of second order, the Kasner exponents of the black hole singularity are continous while the first derivative of the Kasner exponents with respect to the temperature are not continuous. When the phase transition is of first order, the Kasner exponents are not continuous. Our study confirms that the physics of black hole interior is related to the physics outside the black hole horizon. Finally, the Kasner exponents could be obtained from the behavior of the field theoretical observable which are related to the geodesics in the bulk.

There are lots of open questions to be explored. Firstly, since we only considered a particular holographic model with a special choice of parameters, it would be interesting to check how generic our results are. Especially, to check if the behavior of Kasner exponents during the phase transition are universal.  
Secondly, it would be interesting to study the evolution of black hole singularities during the time-dependent dynamical phase transitions, e.g. in \cite{Janik:2017ykj, Bellantuono:2019wbn}. This might help us further understand the possible universal physics of non-equilibrium physics. Thirdly, because the black hole interior changes during the phase transition, it would be interesting to construct the evaporating black hole solutions with phase transitions. In this way, when the black evaporates, the changes of the black hole interior should have nontrivial effect on the Page curve and checking the entropy of the Hawking radiation in this case might help us further understand the physics of the islands proposal \cite{Almheiri:2020cfm}.\footnote{Studies on the Page curve for the eternal hairy black hole can be found in e.g. \cite{Caceres:2021fuw}.} Finally, the physics inside the black hole interior might have some connections to black hole information paradox \cite{Almheiri:2020cfm, Perry:2021mch}. It would be interesting to understand further properties of dynamics of quantum fields close to the black hole singularities and make connections to the physics at the boundary. 

\subsection*{Acknowledgments}
We would like to thank Li Li, Hong Lv, Hua-Jia Wang, Xin-Meng Wu, Run-Qiu Yang, Xin-Yi Zhang for useful discussions. This work is supported by the National Natural Science Foundation of China grant No.11875083. A.R is also supported by Zhuoyue Postdoc Fellowship of Beihang University (ZYBH2018-01).

\vspace{0.95cm}
\appendix
\addtocontents{toc}{\protect\setcounter{tocdepth}{0}}


\section{The boundary terms for the scalar field in holographic renormalization}
\label{app:holoren}
In this appendix we describe a strategy to fix the free parameter $\sigma$ in the counter-term action \eqref{LorRenAction} for the scalar field in holographic renormalization  \cite{Papadimitriou:2007sj,Faulkner:2010gj}. 

Let us first consider the known case of symmetric potential, \emph{i.e.} $\lambda_3=0$. In this case it is known that  $\sigma=0$. The variation of the (renormalized) action \eqref{LorRenAction} with respect to the scalar field $\phi$ is,
\begin{align}\label{eq:varphi}
	\delta S  \supset  \int_{ \mathcal M} d^4x\; \sqrt{-g}\,(\text{EOM})\,\delta \phi + \int_{\partial \mathcal M} d^3x\;\frac{e^{\chi/2}\sqrt{f}}{r^3} \left[ -2 r \sqrt{f}\; \phi\delta \phi' + 2(1+\kappa r)  \phi\delta \phi \right]\,,
\end{align}
where $\partial \mathcal M$ is defined at the boundary  $r=r_c$ with $r_c\rightarrow 0$.  Substituting the expansion \eqref{eq:nbexp} into \eqref{eq:varphi}, it reduces to
\begin{eqnarray}\label{dtboundrycond}
	\delta S \supset \int_{\partial \mathcal M} d^3x \; 2\alpha \,\delta \left(\kappa \alpha-\beta \right)\,.
\end{eqnarray}
The above expression shows that with double trace deformation the source for the scalar operator is $\kappa\alpha-\beta$ while the expectation value is $\alpha$. 

In the non-symmetric potential case, i.e.$\lambda_3 \neq 0$,  we expect  that (\ref{dtboundrycond}) continues to hold.  Now the variation of the action with respect to the scalar filed $\phi$ 
is given by 
\begin{eqnarray}
	\delta S \supset \int_{\partial \mathcal M} d^3x\; \left( 2\alpha \delta (\kappa \alpha -\beta) + 3 \sigma \alpha^2 \delta \alpha + 6 \lambda_3 \alpha^2 \delta \alpha \right)\,.
\end{eqnarray} 
Comparing this to \eqref{dtboundrycond}, we have $\sigma = -2\lambda_3$. 

\end{document}